# Potential habitability of present-day Mars subsurface for terrestrial-like methanogens

**Butturini A.[1], Benaiges-Fernandez R.[2], Fors O.[3], Garcia-Castellanos D.[4]**

1) Departament de Biologia Evolutiva, Ecologia y Ciències Ambientals, Universitat de Barcelona, Diagonal 643, Barcelona, Catalonia 08028, Spain
2) Departament de Genètica, Microbiologia i Estadística, Universitat de Barcelona, Diagonal 643, Barcelona, Catalonia 08028, Spain
3) Departament de Física Quàntica i Astrofísica, Institut de Ciències del Cosmos (ICCUB), Universitat de Barcelona, IEEC-UB, Martí i Franquès 1, 08028 Barcelona, Spain
4) Geosciences Barcelona (GEO3BCN), Consejo Superior de Investigaciones Científicas (CSIC), Barcelona, Spain

Corresponding author: abutturini@ub.edu

## Abstract

The intense debate about the presence of methane in the Martian atmosphere has stimulated the study of methanogenic species adapted to terrestrial habitats that mimic Martian environments. We examine the environmental conditions, energy sources and ecology of terrestrial methanogens thriving in deep crystalline fractures, sub-sea hypersaline lakes and subglacial water bodies considered as analogs of a hypothetical habitable Martian subsurface. We combine this information with recent data on the distribution of buried water/ice and radiogenic elements on Mars and with models of the subsurface thermal regime of this planet to identify a 4.3–8.8 km-deep regolith habitat at the mid-latitude location of Acidalia Planitia, that might fit the requirements for hosting putative Martian methanogens analogous to the methanogenic families Methanosarcinaceae and Methanomicrobiaceae.

## 1) Introduction

One of the main aims of astrobiology is to detect signatures of putative extraterrestrial life (Seager et al., 2012). Methane is one of the most studied biosignatures and its detection, origin and fate is at the center of intense research, particularly on Mars.

The debate over the presence of methane on Mars is long-standing (Sullivan, 1969; Herr and Pimentel, 1969) and continues to be fueled by the controversy surrounding its in-situ detection reported in Gale's crater (Webster et al., 2015), the lack of its detection in the atmosphere using remote sensing techniques (Korablev et al., 2019), and efforts to reconcile the two approaches (Giuranna et al., 2019; Montmessin et al., 2021).

On Earth, Archaea are organisms that release biotic methane as the primary metabolite of their metabolism. Eight types of methanogenic metabolism have been distinguished according to the use of different carbon sources and electron donors (Garcia et al., 2022), the three most relevant here being the following:

a) Hydrogenotrophs, which extract energy from $CO_2$, $HCO_3^-$ or $HCOO^-$ ($e^-$ acceptor) and $H_2$ (the $e^-$ donor).
b) Acetatotrophs, which extract energy from the degradation of acetate.
c) Methyltrophs, which extract energy from the degradation of C1 organic compounds (formate, methanol, and methylamines among others; Kurth et al., 2020).

Methyltrophs can use (or not use) $H_2$ as the $e^-$ donor, adding versatility to their metabolic capabilities (see Appendix 1 for more details about biotic methanogenesis).

The discovery of a simplified ecosystem dominated by a hydrogenotrophic methanogen from the Methanobacteriaceae family in the hot groundwater of Lidy Hot Springs (Chapelle et al., 2022) has prompted speculation regarding the potential existence of a monospecific ecosystem.

Based on this, the methanogenic hydrogenotrophs have emerged as a candidate "model organism" for studying survival on Mars (Boston et al., 1992; Kral et al., 2010; Mickol et al., 2014) because they feed on simple inorganic molecules (i.e. $H_2$ and $CO_2$) that are expected to be available on Mars. Since these molecules can have an abiotic origin this metabolic pathway could operate without the mediating of any other biological metabolism. This conjecture has inspired numerous experiments aiming to test the ability of single methanogenic strains to thrive under the most extreme conditions of temperature, desiccation, salinity, pressure, pH or radiation (see Appendix 2 for an exhaustive list of experimental studies).

Although the findings at Lidy Hot Springs have been criticized (Nealson, 2005; Lin et al., 2006) and, to the best of our knowledge, have not yet been independently validated, it is essential to verify whether the conjecture regarding the existence of methanogenic monospecific communities has solid foundations, or conversely, if it is grounded on an anomalous observation.

The present-day Martian surface and shallow sub-surface is considered hostile for active terrestrial-like organisms (Hallsworth, 2021; but, for a different opinion see Jones, 2018) because of the impact of high-energy radiation (galactic cosmic rays, solar UV, solar energetic proton radiation) and the extremely dry and cold conditions with extreme day–night temperature oscillations when compared to Earth's standards environmental conditions (Atri et al., 2023; Butturini et al., 2022).

In contrast, the Martian subsurface is believed to be more habitable than its surface (Westall et al., 2013; Tarnas et al., 2021) because it is sheltered from the impact of ionizing UV, X-ray and particle radiation. Additionally, subsurface temperatures are higher and exhibit smaller daily and seasonal fluctuations (Jones et al., 2011), likely increasing the probability of storing liquid water (Michalski et al., 2013). Finally, salts are believed to be relevant in the subsurface of Mars (Burt and Knauth, 2003), opening up the possibility of having liquid water at below-zero temperatures. Therefore, the search for a habitable subsurface niche on Mars will involve detecting anoxic habitats with circulating liquid water and available energy and carbon for the growth of Earth-like microorganisms.

According to the literature, three terrestrial subsurface environments are emerging as putative models of the habitable subsurface of Mars:

a) Fluids within deep crystalline-bedrock continental fractures (hereafter DCBF; Sherwood Lollar et al., 2007; Fernández-Remolar et al., 2008).

b) Sub-glacial lentic waters and brines (hereafter SGL; Tung et al., 2005; Mikucki et al., 2015; Forte et al., 2016; Gaidos et al., 2004).

c) Deep-sea hypersaline anoxic basins (hereafter DHAB; La Cono et al., 2019; Fisher et al., 2021).

Of course, DCBF, SGL and DHAB habitats are coarse analogs of the (still poorly-known) Martian subsurface. On Earth, the ubiquitous and oxidant biosphere fueled by the Sun's power has challenged the search for an optimal terrestrial analog of the anoxic Mars subsurface. An "ideal" terrestrial analog should harbor a self-sufficient ecosystem with minimal hydrologic, geologic, atmospheric and biological interactions with the Earth's surface (Lollar et al., 2021). But, being realistic, only a few DCBF sites are likely to roughly fulfill these requirements for remoteness (Lin et al., 2006; Holland et al., 2013). Further, most SGL and DHAB sites studied to date are not isolated from surface processes (see Appendix 3 for more details).

Being aware of these limitations, SGL sites provide information about microbiomes at temperatures significantly lower than those reported at DCBF sites. On the other hand, microbiomes from DHAB sites (together with some SGL sites) thrive under conditions of extreme salinity. An example of such cold and saline waters includes the debated subglacial liquid waters in the Mars South Pole, modeled on the basis of radar profiles (Orosei et al., 2018) and anomalies in surface topography, (Arnold et al., 2022) although alternative explanation of radar and topography data exists (Lalich et al., 2024 and Sori et al., 2024).

To sum up, the topic of Mars's habitability has unified various field studies focused on DCBF, SGL, and DHAB habitats, highlighting the need to integrate this wealth of information. Within this framework, insights gained from these terrestrial environments, and a deeper understanding of the Martian subsurface, are crucial for addressing the following questions regarding potential life on Mars:

1. Are methanogens significant in DCBF, SGL and DHAB terrestrial habitats?
2. Do the intrinsic environmental conditions of DCBF, SGL and DHAB habitats shape the significance of subsurface methanogens and their metabolism?
3. Is the almost "single species" methanogenic ecosystem reported at Lidy Hot Springs representative for DCBF, SGL and DHAB systems?
4. How do the answers to the previous questions impact on the viability of terrestrial methanogens in the subsurface of Mars?
5. What environments in Mars's subsurface could support terrestrial-like methanogens? What methanogenic pathways may prevail?

To deal with these questions, the present study followed two approaches: first, we compiled information about the presence of methanogens in DCBF, SGL and DHAB systems (Section 3). Then, we reviewed recent advances on the geology of the Martian subsurface to identify a region and subsurface that could host terrestrial-like methanogens (Section 4).

Several authors have reviewed separately the significance of microbiota in DCBF, SGL and DHAB (Kieft, 2016; Magnabosco et al., 2018; Soares et al., 2023; Merlino et al., 2018; Varrella et al., 2020; Fisher et al., 2021; Achberger et al., 2017). However, to date, no study has compared the composition of microbial communities across these three habitats in the context of Mars's habitability. Therefore, we specifically review which methanogens have been reported in terrestrial subsurface ecosystems (Section 3.1), whether they coexist with other organisms (Section 3.2), and which energy sources they consume (Section 3.3).

To synthesize recent research advances on the geophysics of Mars's lithosphere we focus on: a) information about the availability of subsurface water or ice (Section 4.1); b) information providing clues about the possibility of producing abiotic molecular hydrogen, the simplest and most widely used energy source for methanogens (Section 4.2); and c) how subsurface temperature increases with depth until it becomes tolerable for terrestrial-like methanogens (Section 4.3). Thus, we focus on water, energy sources and temperature. In contrast, the availability of inorganic carbon, the main electron acceptor for methanogens, is not discussed. Recognizing the limited information about inorganic carbon in Mars's subsurface (Ehlmann and Edwards, 2014; Wray et al., 2016; Jakosky, 2018), in this study we assume that carbon is not a limiting factor.

Combining information from these two approaches provides the basis with which it is possible to determine more accurately the location, extent and depth of Mars's subsurface lithosphere that is potentially habitable for terrestrial-like methanogens and to identify which methanogenic metabolism would adapt best to these putative habitats.

## 2) Data & Methods

This study combines the information provided by research focused on the microbiological and environmental characteristics of DCBF, DHAB and SGL habitats as well as the geological, geochemical and hydrological characterization of Mars's subsurface. A summary of the Methods is provided below. Additional details are in Appendix 4.

2a) *Microbial data source*

The dataset included microbiological and environmental data from 79 sites (49 DCBF, 11 SGL and 19 DHAB sites). The criteria for site selection are described in Appendix 4. The compiled dataset and bibliographic sources are in Appendices 5 and 6.

2b) *Salinity proxy*

Being aware of the expected relevance of salinity in Martian environments (Burt and Knauth, 2003), the molar ionic strength (I) of fluids was calculated. Appendix 4 reports the method used to estimate the ionic strength at the selected study sites.

2c) *Mars's subsurface habitats*

Information about the presence of water, the availability of energy sources (for putative methanogens) and the geothermal gradients of Mars's subsurface was collected from the literature.

*Water/ice availability:* Information about water or ice in Mars's subsurface originates from in-situ and remote measurements such as: radar profiles obtained by the Zhurong rover; seismic data from the upper crust obtained by the InSight lander; images of the surface at the Phoenix and Viking 2 landing sites; and remote-sensing data provided by the Mars Odyssey Mission Gamma Ray Spectrometer (GRS), the Mars Advanced Radar for Subsurface and Ionospheric Sounding instrument (MARSIS), and the Mars Climate Sounder (MCS). Morgan et al. (2021) mapped the information obtained from multiple remote-sensing tools into contingency maps ($Cv_{(z,\ long,lat)}$) that indicate the likelihood of the presence of ice at the surface (0–1 m depth, $Cv_{(0, long,lat)}$) and of shallow buried ice (>5 m depth, $Cv_{(5 long,lat)}$) from 60ºS to 60ºN.

*Energy sources for methanogens:* The focus is on the viability of abiotic production of $H_2$ because it is a crucial energy source for terrestrial methanogens. Four main mechanisms are proposed to promote $H_2$ accumulation in the Martian subsurface: i) $H_2$ diffusion from the atmosphere to the subsurface dry regolith; ii) rock crushing after seismic movements; iii) water–mineral interactions such as serpentinization within crustal fractures and iv) water radiolysis driven by radiogenic Heat-Producing Elements (HPE: $^{232}Th$, $^{40}K$ and U).

Information related to *i)* and *ii)* originates from theoretical considerations and models. The significance of items iii) and iv) has been explored by combining information provided by the Compact Reconnaissance Imaging Spectrometer (CRISM) and the GRS instruments with modelization. Among the HPE, $^{232}Th$ is the most important heat source and its concentration ($[Th]_{(long,lat)}$) on Mars's surface has been mapped across all longitudes and from 60ºS to 60ºN (Hahn et al., 2011).

*Geothermal gradients:* Temperature gradients in the subsurface Martian upper crust can be estimated using the Fourier Law and whatever information about average temperature and heat flow at the surface exists, together with clues about the thickness, densities and thermal conductivities of the geologic structures expected in the Martian upper crust. This information was compiled from the literature and is reported in Appendix 4. The geothermal gradients were modeled in the region with the highest likelihood of hosting subsurface water and producing $H_2$ together with the highest average surface temperature to locate the shallowest subsurface liquid water datum. At the target site, we estimated the depths at which temperatures are expected to be viable for methanogens according to the information collected from DCBF, SGL and DHAB sites.

## 3) Methanogens from terrestrial subsurface habitats

### 3.1) Methanogens in DHBA, DCBF and SGL

Methanogens were reported at 66% of sites and in 21% of cases were considered a relevant component of the microbiome. Methanogens are ubiquitous in DHABs but were not reported at 45% of DCBF and SGL sites.

Regardless of their ecological significance, methanogens were reported to have a low relative abundance (less than <1% in most cases; Lavalleur and Colwell, 2013; Lopez-Fernandez et al., 2018; Nuppunen-Puputti et al., 2018; Achberger et al., 2016; Kalwasińska et al., 2020).

Methanobacteriaceae and Methanosarcinaceae were the most frequently reported families at DCBF and DHAB sites. At SGL sites Methanomicrobiaceae and Methanosarcinaceae predominated over other families (Figure 1a).

Hydrogenotrophics and methylotrophics were more frequently reported than acetatotrophics at DCBF and DHAB sites, respectively. At SGL sites, the three methanogenic pathways were of similar significance (Figure 1b).

Knowledge on rates of methanogenesis is scarce. Therefore, we know little about methanogenic activity at the study sites. Despite the small amount of information available, the rate of methanogenesis appears to be higher at DHBA sites (in the range 0.5–85 μM/d) than at DCBF (<2.2 μM/d) and SGL sites (<0.45μM/d; see Appendix 6 for more information).

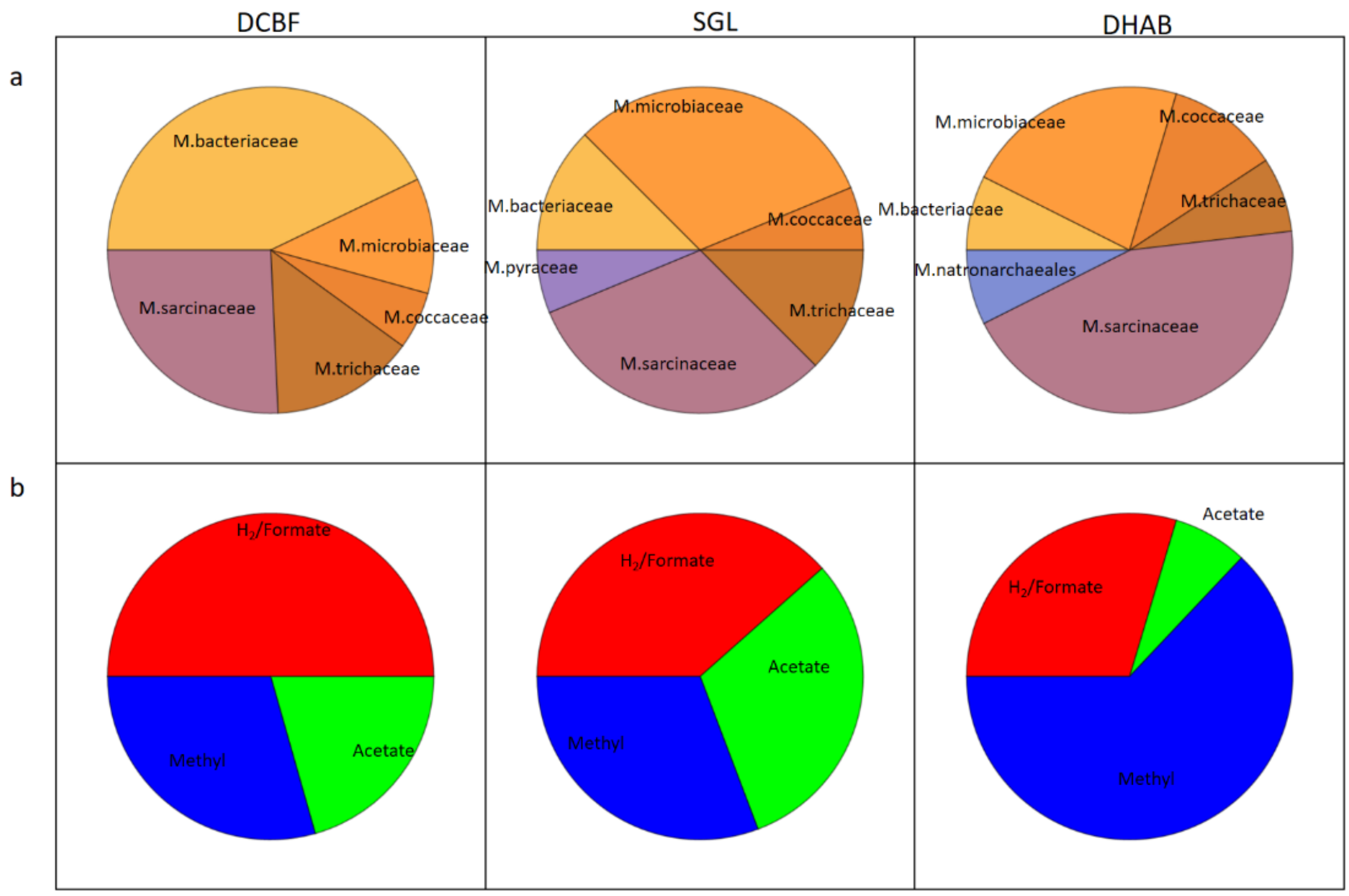

Figure 1. The pie charts show the percentage at which each family of methanogens (panel a) or each methanogenic pathway (panel b) has been reported in the three selected habitats (DCBF: deep crystalline-bedrock fractures fluids; SGL: Sub-glacial lentic waters; DHAB: Deep-sea hypersaline anoxic basins).

Methanogens have been reported in the temperature range -17.5–82°C (Figure 2a), the Methanosarcineceae and Methanomicrobiaceae being the most flexible family and most frequently reported in cold waters (<15°C). In contrast, Methanobacteriaceae are typically reported in temperate-hot (T>15°C) fluids. The most significant exception is Tarn Flat interglacial organic-rich brine (temperature has not been reported but is likely below zero degrees Celsius), which hosts a highly diverse methanogenic community including Methanobacteriaceae (Papale et al., 2019).

Methanogens have been reported to be active up to salinities of $I \sim 9$ M (Yakimov et al., 2013; Borin et al., 2009; Guan et al., 2015). In contrast, Telling et al. (2018) reported a maximum salinity tolerance threshold for the active microbiome in DCBF of $I \sim 2$ M (Figure 3).

DHBAs are the most hypersaline subsurface systems, yet brines accumulated at some SGL sites (Lakes Vida, Vanda and Bonney and endoglacial Antarctic brines of Northern Victoria land and Tarn Flat) and at some DCBF sites (Kidd Creek, Thompson mine and Soudan iron mine). Methanosarcinaceae is the most salt-tolerant family. Methanobacteriaceae, Methanococcaceae and Methanotrichaceae are usually linked to fresh and sub-saline habitats (i.e. $I < 2$ M). Methanomicrobiaceae have been reported in brines with an ionic strength up to 4.5 M (Figure 2b).

Methylotrophs typically predominate in brines with $I > 4$ M, whereas the hydrogenotrophs preferentially thrive under low salinities (Figure 2b). Nevertheless, exceptions exist and the below-zero temperature brines in the Northern Victoria Land endoglacial SGL host a strictly hydrogenotrophic Methanomicrobiaceae (Guglielmin et al., 2023).

The impact of sulfate overlaps with that of salinity: Methanobacteriaceae are typically constrained in water with $[SO_4^{2-}] < 2$ mM. In contrast, Methanosarcinaceae are much more tolerant and are reported to survive in up to 0.4 M of sulfate (Figure 2c).

Methanogens have been reported in the pH range from 4.5 to 11.3. Methanosarcinaceae and Methanomicrobiaceae are preferentially reported in neutral and acid fluids whereas Methanobacteriaceae have been reported in water at $pH > 6.5$ (up to $pH \sim 11.3$ at Samail Ophiolite; Rempfert et al., 2017) (Figure 2d). Hyperalkaline fluids host methanogens at the Samail Ophiolite site (Kraus et al., 2021; Nothaft et al., 2021; Fones et al., 2019) but have not been reported at the Coast Range Ophiolite (Twing et al. 2017) or Cabeço de Vide (Tiago and Vérissimo, 2013) sites.

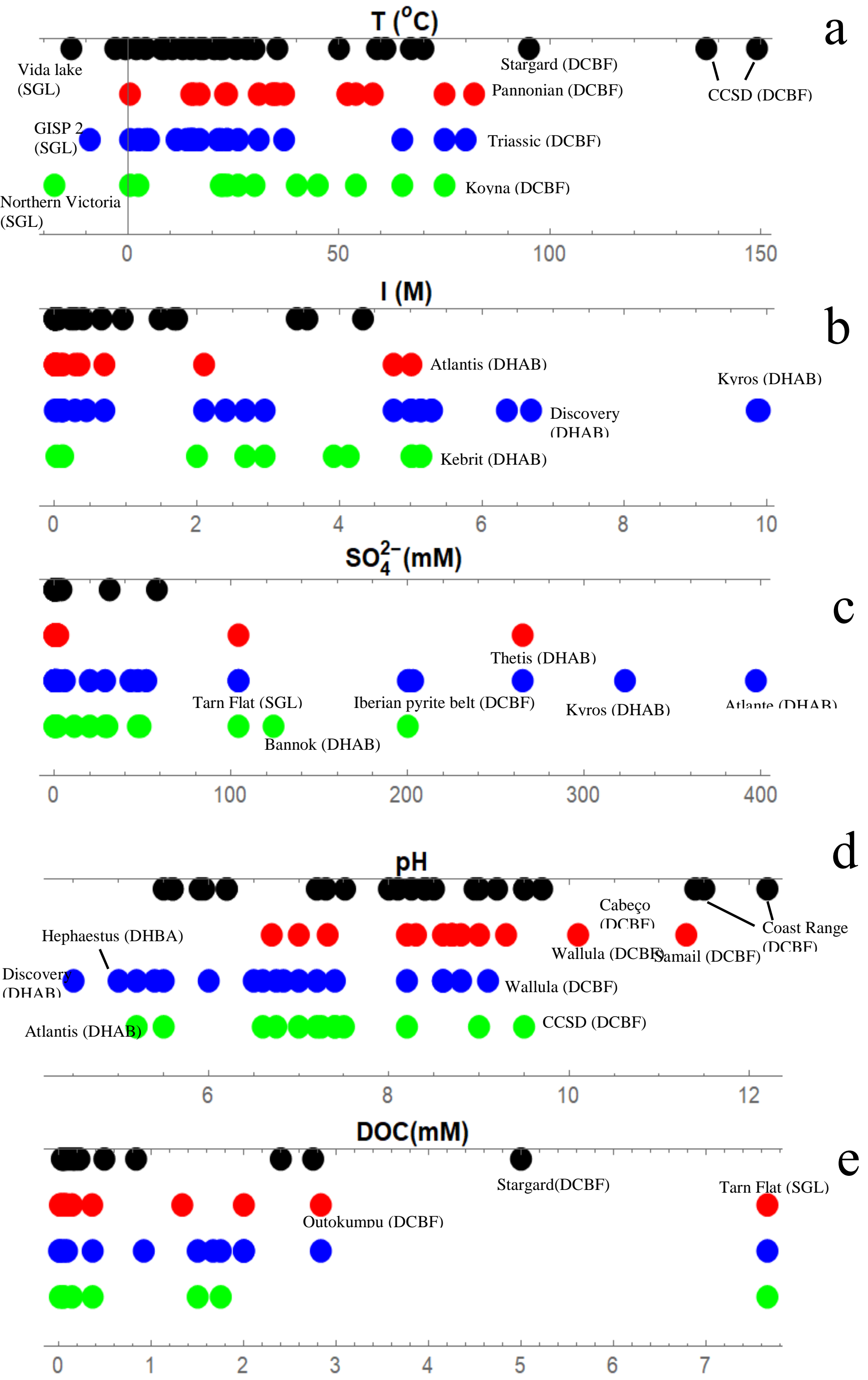

Figure 2. Range of environmental conditions reported at the compiled DCBF, SGL and DHAB sites. Colors indicate the family detected at each site (only the three most frequent

families are shown): Red dots=Methanobactericaeae; Blue dots=Methanosarcinaceae; Green dots=Methanomicrobiaceae; Black dots=Not detected. The names of a few sites are provided. The data used to generate the plots are in Appendix 5.

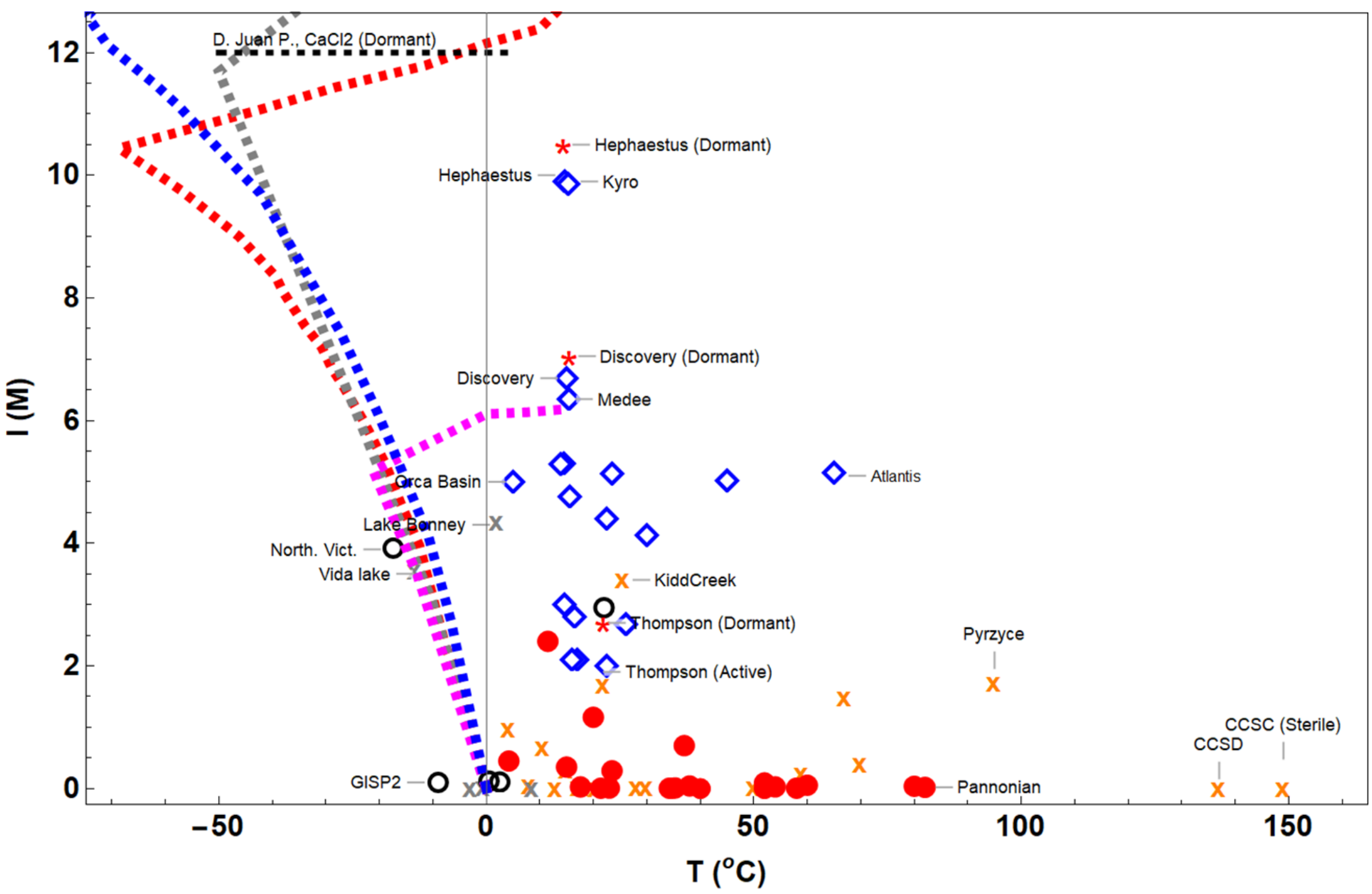

Figure 3. Relationship between salinity (in terms of ionic strength) versus temperature at the study sites. DCBF: red dots; SGL: open black circles; DHBA: blue rhombuses. Crosses indicate sites where the presence of methanogens has not been reported. The names of a few emblematic sites are shown. Dotted lines show the phase diagrams of NaCl (magenta dashed line), $Mg(ClO_4)_2$ (red dashed line), $Ca(ClO_4)_2$ (blue dashed line), and $CaCl_2$ (gray dashed line). The horizontal segmented black line refers to the range of temperatures reported in the hypersaline Don Juan Pond in Antarctica (a $CaCl_2$ brine apparently devoid of active life in the McMurdo dry valley; Oren, 2013).

3.2 Methanogens and other organisms.

With the exception of the Lidy Hot Springs site (Chapelle et al., 2022), methanogens, when present, coexist with other organisms. DCBF, SGL and DHAB host taxonomic and metabolically diverse microbial communities that include bacteria, archaea, viruses (Miettinen et al., 2015; Mikucki et al., 2016; Nuppunen-Puputti et al., 2021; Holmfeldt et al., 2021) and even eukaryotes (Guglielmin et al., 2023; Epova et al., 2022). Bacteria typically predominate over archaea in most habitats (Magnabosco et al., 2018; Guan et al., 2015; Murray et al., 2012). Yet, in addition to Lidy Hot Springs, a notable abundance of archaea (always coexisting with bacteria) has been reported in Olkiluoto (Bomberg et al., 2016), the Tau Tona gold mine (South Africa; Magnabosco et al., 2016) and in some Mediterranean deep-sea brines (Urania, Bannock and Atlante; van der Wielen et al., 2005).

Besides methanogenesis a plethora of autotrophic and heterotrophic metabolic pathways has been reported (for a list see Appendix 7). The core metabolisms are probably those linked to sulfur cycling (Telling et al., 2018; Lollar et al., 2019; van der Wielen et al., 2005).

Methanogens coexist with organisms that, a priori, compete for the same resources (mainly $H_2$). Thus, methanogenic hydrogenotrophs are not outcompeted by sulfate/sulfur autotrophic reducers (Lavalleur et al., 2013; Magnabosco et al., 2016; Lin et al., 2006; Chiriac et al., 2018; Sanz et al., 2021; Plugge et al., 2011; Puente Sánchez et al., 2014) and do not exclude the bacterial homoacetogens (Rempfert et al., 2017; Lopez-Fernandez et al., 2018; Kotsyurbenko et al., 2001). Finally, syntrotrophic bacteria-methanogen consortia are frequently reported at DHBA (La Cono et al., 2015), DCBF (Moser et al., 2005; Osburn et al., 2014; Lau et al., 2016, Sheik et al., 2021) and SGL (Papale et al., 2019) sites.

### 3.3) Energetic resources for methanogens

Information about $H_2$ was available for 29 sites. In most cases, the concentration was higher than 1 nM, the minimal reported $H_2$ concentration that enables hydrogenotropic methanogenesis (Heuer et al., 2009). Most of these sites are DCBF sites, except for the two DHBA sites in the Gulf of Mexico and one SGL site (Lake Vida). Methanogens were not reported at 9 of these 29 sites (Appendix 4). As well as hydrogen, concentrations of formate and acetate were reported at 22 sites (Appendix 6). Methanogens were not reported at 6 of these 22 sites. Therefore, the presence of $H_2$, formate or acetate does not guarantee the proliferation of methanogens.

At DCBF sites, the presence of $H_2$ is attributed to water–rock interactions (serpentinization and rock friction) and water radiolysis. Information about the availability of $H_2$ is lacking for most DHBA sites. However, its availability is plausible at those DHBA sites in contact with deep gas venting such as in the Gulf of Mexico brines and some Red Sea brines (Merlino et al., 2018). Information about $H_2$ at SGL sites is scarce. However, silicate–water laboratory experiments suggest that abiotic $H_2$ production in glacial habitats might be significant (Samarkin et al., 2010; Telling at al., 2015; Gill Olivas, 2019).

The detection of the hydrogentrophic Methanomicrobiaceae in the SGL endoglacial brines indicates that $H_2$ should be available. The Mg/K ratios in these brines and seawater are identical (Guglielmin et al., 2023), suggesting minimal rock weathering. Thus, the production of abiotic $H_2$ through water–rock fractioning appears improbable. $H_2$ is likely of biotic origin from the fermentation of organic matter and acetate oxidation. Indeed, the concentration of dissolved organic carbon is extremely high at these SGL sites (up to 7 mM; Papale et al., 2019, Figure 2e).

The occurrence of acetate and formate at deep DCBF sites has been attributed to abiotic processes related to the reduction of dissolved inorganic carbon with $H_2$ previously released after water radiolysis (Lollar et al., 2021) or serpentinization (Fones et al., 2019). Water radiolysis also generates strong oxidants such as OH•, $O_2$, $H_2O_2$, and $O^-$, yet the impact of these agents on the endolithic biota remains unknown. Further, these oxidants can react with reduced elements from minerals (like pyrite) or with solutes (like $NH_4^+$) and produce $SO_4^{2-}$ , $Fe^{+3}$ or $NO_2^-/NO_3^-$ that can fuel other autotrophic metabolisms (Lefticariu et al., 2010; Silver et al., 2012), which, in turn, might compete with methanogens for $H_2$.

## 4. Potential habitats in Mars's subsurface

### 4.1 Water/ice availability

Phoenix and Viking 2 revealed that a very shallow (few centimeters depth) subsurface ice is widespread and stable at latitudes higher than 45°N of the northern hemisphere (Mellon et al., 2009 and 2022) that likely extends to several meters in depth (Putzig et al., 2014). In contrast, Zhurong and InSight indicated that, at low latitudes (<25° N) of the lowlands, the shallow and deep crustal regolith is apparently dry with weak evidence of the presence of ice or liquid water (Manga & Wright, 2021; Li et al., 2022, Wright et al., 2022). Maps of the likelihood of ice (Morgan et al., 2021) show the widespread presence of near-surface ice at latitudes higher than 60 degrees. However, buried cryosphere (>5m depth) is relatively likely even at latitudes down to 30°N in the northern hemisphere (Putzig et al., 2023, Figure 3a). Similarly, Piqueux et al. (2019) estimated that shallow buried ice (at less than 1 m depth) is widespread at latitudes as low as 40°N. Finally, interpretation of MARSIS echoes has revealed the location of putative subglacial layered deep brine deposits at the South Pole (Lauro et al., 2022; but for a different interpretation of MARSIS echoes see Lalich et al., 2024) and putative deep ice-rich deposits in the Medusae Fossae Formation, at the Martian equator, below a 300–600 m-thick dry upper regolith (Watters et al., 2024). This latter discovery suggests that there might be subsurface ice at the present-day Mars equator, diverging somewhat from conclusions based on InSight data.

4.2 Availability of energy sources

In the present-day Martian atmosphere, molecular hydrogen can be released after photochemical reactions. Sholes et al. (2019) modeled the possibility of passive diffusion of atmospheric $H_2$ through the regolith and the possibility that it could sustain hydrogenotrophic methanogenesis in a subsurface connected to the atmosphere. A priori, this mechanism should be effective at low latitudes (<30º) where the upper subsurface is highly likely to be dry (i.e. ice-free, Figure 4a), allowing the atmospheric gases to diffuse through the regolith interstices. Despite the existence of this theoretical approach, empirical evidence that provides clues as to the relevance of this process remains elusive, to the best of our knowledge. This leads to the conclusion that, even if the diffusion of photo-produced $H_2$ in the atmosphere turns out to be possible, it would occur only in subsurface habitats that are intrinsically uninhabitable because they do not contain water. On Earth, friction, fracturing and crushing of hydrated minerals or the reaction of silicon radicals with water (Kita et al., 1982) as a consequence of seismic movements (or meteoritic impacts) promote the liberation of seismogenic hydrogen. McMahon et al. (2016) hypothesized that seismogenic hydrogen in the Martian upper crust could sustain hydrogenotrophic methanogenesis. This mechanism is expected to be activated by seismic events. The InSight mission demostrated that Mars is seismically active, especially at Cerberus Fossae and along or north of the dichotomy between the southern highlands and northern lowlands, especially at the Tharsis plateau (Ceylan et al., 2023).

However, to date, there is a large uncertainty about the presence of subsurface water/ice in these regions (Burr et al, 2002; Morgan et al., 2021, Figure 3a). Consequently, the current information does not strengthen the feasibility of $H_2$ production from Martian present-day seismic activity.

Serpentinization is a low temperature water–rock process that produces hydrogen on Earth's crust. Amador et al. (2018) concluded that clues indicating serpentine systems are rare on Mars's surface. Indeed, it is only in the eastern portion of Nili Fossae that the detection of serpentine together with other related minerals (talc and saponite) clearly coincides with the detection of a high concentration of olivine-rich basalt outcrops, making past serpentinization in this area plausible. Although water–rock processes in the present-day Martian crust cannot be ruled out (Vance et al., 2020), information provided by CRISM does not support globally widespread serpentinization in the early Mars when the planet was geologically more active, and does not provide any indication of active serpentinization on present-day Mars. In sum, the available information suggests that a significant $H_2$ yield from serpentinization in the present-day Mars subsurface is unlikely.

In a water/ice-rich subsurface, radiogenic elements (HPE) fuel water radiolysis with the release of $H_2$ (among other molecules) that might support a hydrogenotrophic community (Dzaugis et al., 2018). According to information provided by GRS (Hahn et al., 2011), the concentration of Th on Mars's surface is approximately one-tenth of that reported in the terrestrial crust (Lin et al., 2005) but is in the same range as that reported in subsea basalts (Dzaugis et al., 2016).

The HPE concentration in the subsurface is unknown, but seismic data from InSight suggest that the Martian crust is enriched in radiogenic elements (Knapmeyer-Endrun et al., 2021). Assuming that the abundance of radiogenic elements in the subsurface is the same as that detected at the surface (Hahn et al., 2011), the radiogenic heat production of the present-day Mars crust would range between 2.5 and 7.5 x $10^{-5}$ μW $kg^{-1}$ and would correlate with the abundance of Th, the largest heat source (Hahn et al., 2011). If interstitial water is in contact with such a radiogenic flux, $H_2$ release is plausible (Dzaugis et al., 2018).

The size of the void volume in sediments and rock fractures influences the efficiency of water radiolysis. Indeed, $H_2$ production is expected to be higher in low-porosity sediments (<5%) and small fractures (<1 μM width; Dzaugis et al., 2018). It is expected that the porosity of the first 4 km of the Martian crust could range between 10 and 63%, with the highest values corresponding to the upper centimeters of the regolith (Clifford et al., 2010; Goossens et al., 2017; Michalski et al., 2018; Kilburn et al., 2022; Grott et al., 2021). Thus, although the porosity of the Martian regolith of the upper crust is currently poorly constrained, radiolytic $H_2$ production is likely to be more efficient in deeper environments with lower porosity and fracturing compared to the shallow subsurface.

According to Hahn et al. (2011) there are four regions on Mars's surface with a Th abundance higher than the 95th percentile (Figure 4a, labeled A to D). Areas A (the largest one) and B are at mid-high latitudes (from 30 to 60°N of the northern hemisphere). Area C (Hephaestus Rupes) is at a lower latitude (20 °N) close to the Zhurong landing site. Finally, area D coincides with the Eridania region, a zone that could have hosted a radiogenic hydrothermal system in the Noachian (Ojha et al., 2021).

We intersected the map of the ice contingency map at a depth >5 m $Cv_{(5,long.lat)}$ (Morgan et al., 2021) with that of the Th abundance ($[Th]_{(long,lat)}$) to identify regions with the highest chance of promoting water radiolysis in the present-day Mars subsurface. Therefore, this intersection helps to identify regions with the highest likelihood of hosting subsurface terrestrial-like methanogens. Locations at which $Cv_{(5\ long,lat)}$ and $[Th]_{(long,lat)}$ values are higher than the 95th percentile are selected. This intersection reveals that the areas labeled as A and B in Fig. 4 have the highest probability to store a subsurface buried/shallow water/ice layer together with a high availability of HPE. In contrast, areas C and D were removed from the model because evidence of the presence of subsurface water/ice is currently weak (Figure 4b). It is noteworthy that, according to planetary-scale crustal models (Parro et al., 2017), areas A and B are likely to experience low heatflow values, both at the surface and the base of the crust, implying a deeper depth for the 0 ºC isotherm.

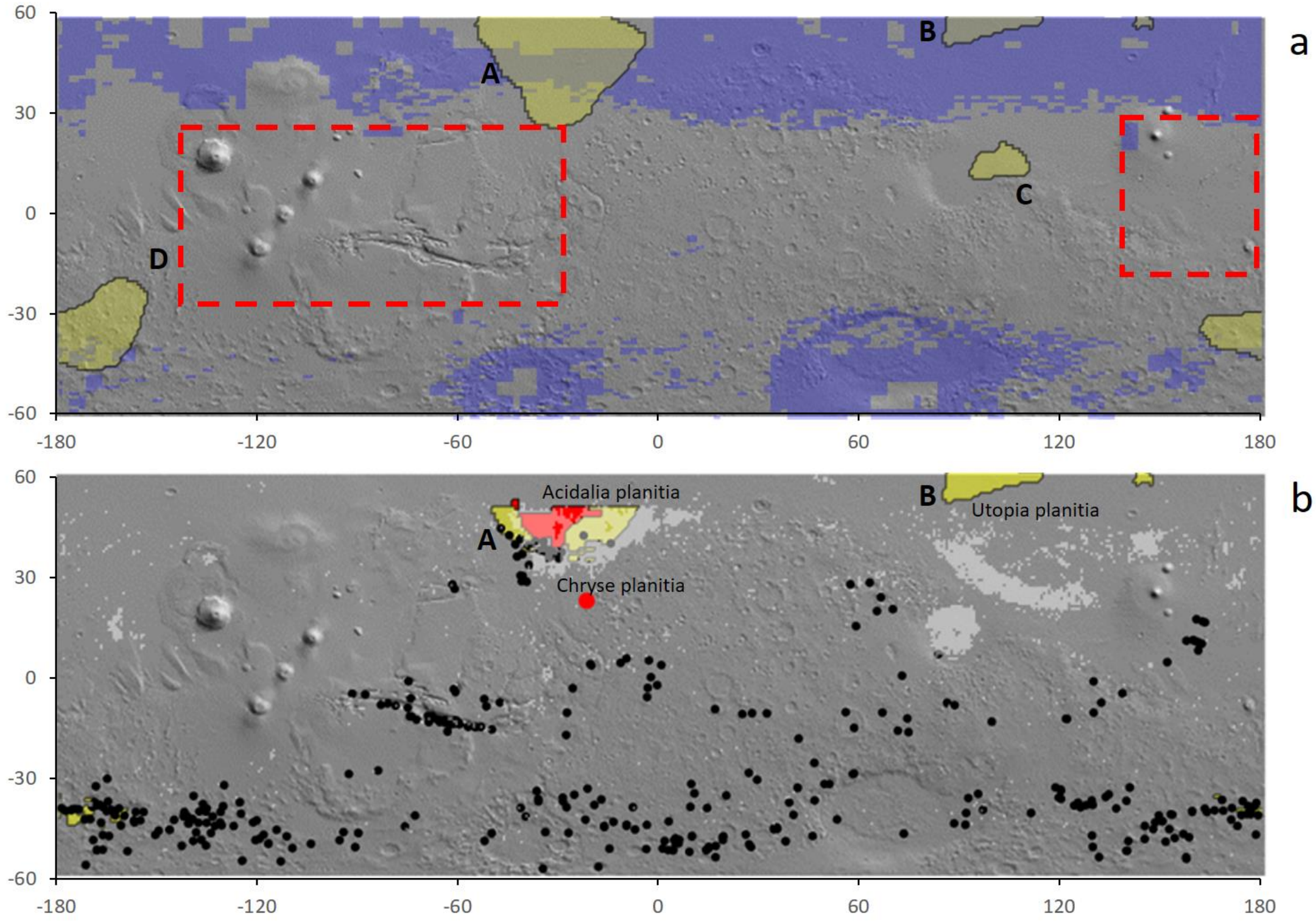

Figure 4. Panel a: spatial distribution of 95th percentile values of ice contingency (at depth >5 m; light blue shaded area) and of Th concentration (yellow shaded area labeled with capital letters). Dashed red rectangles delimit the most seismically active regions according to the InSight mission (Ceylan et al., 2023). Panel b: intersection of the two previous maps. Yellow and red shaded areas (labeled with A and B) show the regions where values of both the 95th percentile and 99th percentile of $Cv_{(5,\ long,lat)}$ and $[Th]_{(long,lat)}$ coincide. The faint white dots show the spatial distribution of pitted cones structures (modified from Mills et al., 2024). The black dots show the spatial distribution of Slopes

Recurrent Lineas (SRLs) according to Mcewen et al., (2021); Stilmann et al., (2016), and Oijha et al., (2014). The red dot shows the McLaughlin Crater with exposed carbonate rocks and signs of groundwater activity (Michalski et al., 2013). Gray in the background shows the Mars topography based on MOLA.

4.3 Subsurface thermal regime

The last step involves calculating the geothermal gradient at one location within the selected areas A and B to find the depth at which the minimum temperature viable for terrestrial methanogens (i.e. -17.5 C) and the melting temperature of ice are exceeded.

Within the two areas, the most robust location is in the southern Acidalia Planitia (35°N, 30°W, red area at Figure 4b), in contact with Chryse Planitia, with an annual average temperature of -58°C (according to the Mars climate database). This average temperature is higher than that estimated at the rest of the sites located inside areas A and B, because it is at the lowest latitude.

According to Parro et al. (2017), the heat flow at the surface of this location is approximately 0.016 W $m^{-2}$. To constrain the subsurface geology and thermal conductivities, we adopted the information derived by the Zhurong rover and especially by InSight lander. The Zhurong landing site and the Acidalia Planitia are within the limits of a putative ancient ocean in the northern hemisphere (Di Achille & Hynek, 2010), in the same geological unit (Late Hesperian lowlands unit, Tanaka et al., 2014), with a low dielectric constant indicating the presence of low-density sediments or deposits of ground ice or both (Mouginot et al., 2012).

Radar reflections detected by Zhurong allow the identification of three main structural sedimentary layers below the landing site (Li et al., 2022): a regolith top layer probably of fine-grained sediments of approximately 10 m in depth; a second layer that extends up to 30 m in depth, mainly composed of unevenly distributed small boulders/cobbles; and a third layer that extends down to 80 m below the surface where large, clast-supported boulders are distributed more evenly. The deeper subsurface layers remain unconstrained at the Zhurong site. However, the InSight seismograph detected structural discontinuities at 2, 8–11 and 20 km below the surface (Shi et al., 2023; Knapmeyer-Endrun et al., 2021). The 2-km deep discontinuity has been interpreted as a structural transition between a highly fractured basaltic layer (called a "megaregolith") and a more coherent and less fractured layer (Shi et al., 2023; Carrasco et al., 2023). The discontinuity detected at 8–11km deep has been attributed to the presence of rigid flood basalt (Xu et al., 2023) putatively constituted by an unfractured plagioclase feldspar rock (Kilburn et al., 2022). The boundary between the crust and the uppermost mantle is between 20 and 39 km (Schimmel et al., 2021; Knapmeyer-Endrun et al., 2021). Importantly, the InSight seismograph indicated the existence of a poorly-cemented upper crust (Wright et al., 2022), suggesting that the significance of salts at subsurface at low and mid latitudes of northern lowlands is doubtful. Notably, salts also seem to be in low abundance in the Jezero crater (Wiens et al., 2022), which is also located at low latitudes in the northern hemisphere (18.4°N; 77.6°E).

The thermal conductivity of the dry Martian upper low-density regolith is constrained between 0.085 and 0.039 W/m/K (Zent et al., 2010; Grott et al., 2021). Li et al. (2022) modeled the temperature profile down to a depth of 140 m at the Zhurong site using a constant thermal conductivity of 0.8 W/m/K (as suggested by Egea-González et al., 2021) regardless of the lithological vertical heterogeneity described above. Instead, to model the steady-state temperature profiles in the Acidalia Planitia subsurface, we changed the substrate's electrical conductivity according to its saturation level (dry or saturated), using values from the literature tabulated in Table S4.. This was modeled using the Fourier

Law, assuming a subsurface with two main layers: an upper 2 km-thick layer of highly fractured megaregolith with moderate density (1700 $kg/m^3$; Parro et al., 2017) and a bottom 7 km-thick denser layer (2850 $kg/m^3$; Parro et al., 2017) of fractured basalt ice-saturated upper crust. A thin, shallow upper layer of less than 10 m consisting of high porosity and low thermal conductivity substrate was not included in the calculations because its small thickness makes it irrelevant to the temperature results.

Temperature profiles stop at a depth of 9 km where seismic data from the InSight suggests a structural discontinuity that could reflect the presence of an unfractured (zero-porosity) flood basalt (Kilburn et al., 2022; Xu et al., 2023). Figure 5 shows the three subsurface temperature profiles estimated based on the water/ice boundary being at z=150, 1000, and 2000 m depth. In the first case, T=0°C is reached at a depth of 8.8 km, while in the second and third scenarios ice melts at 6.8 and 4.3 km depth, respectively. The melting depths would decrease to 5.0, 3.2, and 1.8 km respectively if water is NaCl-saturated. However, as mentioned before, the low abundance of cement minerals inferred by InSight makes this scenario unlikely or perhaps constrained to small and isolated salty lenses.

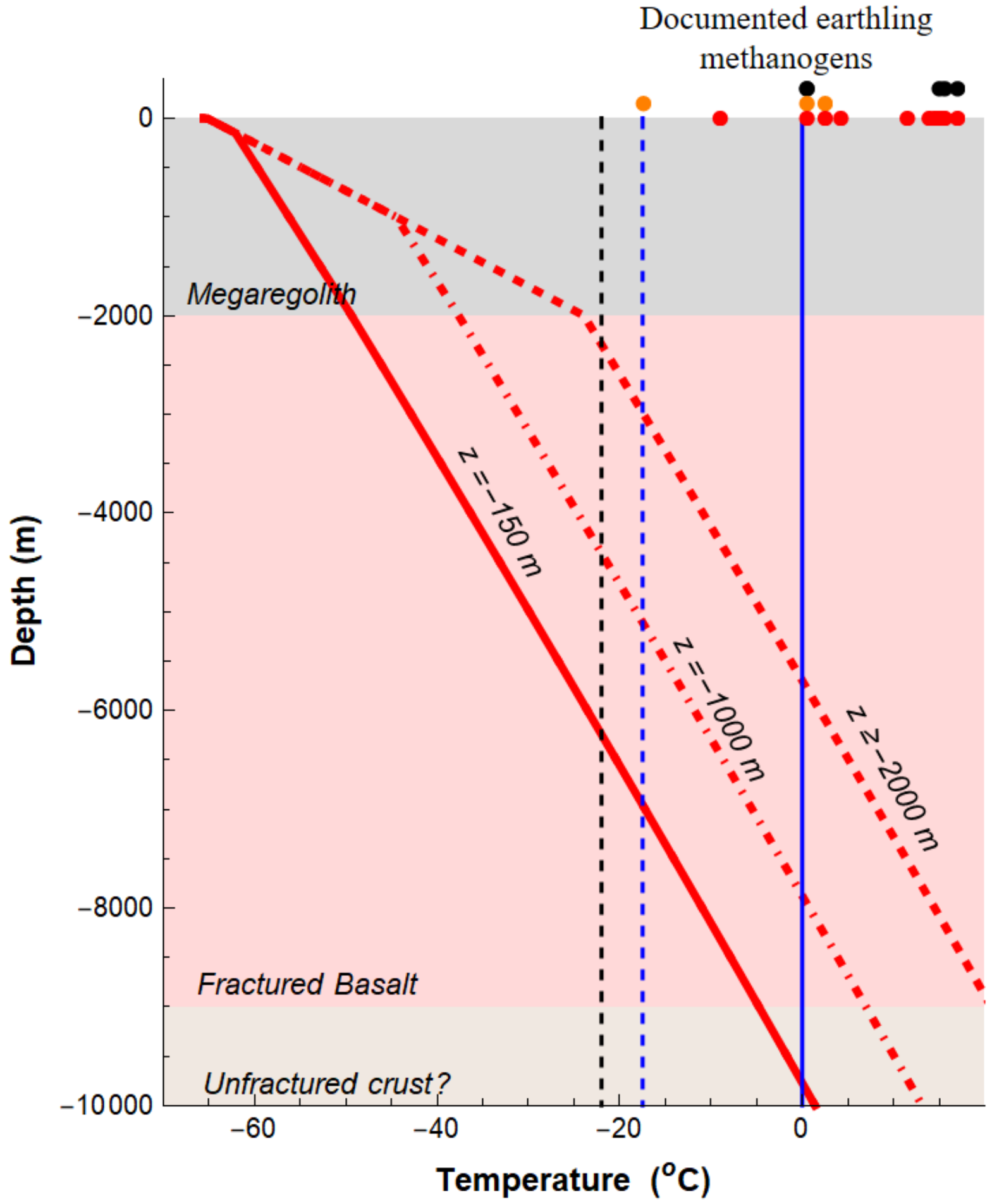

Figure 5. Subsurface temperature profiles (red lines) at 35N, 30W (south of Acidalia Planitia) modeled based on steady-state heat flow using a yearly-average surface temperature of –58 C. The three geotherms correspond to three depths *z* with no ice above them, reducing thermal conductivity in the upper-most kilometers. Vertical lines show the fusion temperature for freshwater water (0°C; blue line), NaCl eutectic temperature (-21°C; dashed black line) and the lowest temperature reportedly tolerated by terrestrial underground methanogens (-17.5°C; Guglielmin et al., 2023; dashed blue line). The top x-axis shows the temperatures at which terrestrial subsurface methanogens are reported in the subsurface, as described in Section 3. The yellow, orange and red dots are the Methanobacteriaceae, Methanomicrobiaceae and Methanosarcinaceae families respectively.

5) Discussion

The results provide insights into the questions raised in the introduction:

1) Are methanogens significant in DCBF, SGL and DHAB habitats?

Methanogens are significant in terrestrial subsurface habitats at several study sites. However, they are not ubiquitous and should not be regarded as core members of the terrestrial subsurface ecosystems. Indeed, some DCBF sites that apparently fulfill the conditions for methanogenesis (anoxic fluids, at moderate-high temperature, abundant $H_2$, formate, acetate) do not show unquestionable evidence that they host methanogens. Thus, the proliferation of methanogens in terrestrial subsurface habitats is a response to intricate interactions with the physical environment and with organisms possessing different metabolisms. $H_2$ is the crucial electron donor for many other anaerobic metabolisms reported in terrestrial subsurface ecosystems. Indeed, under standard conditions the bioenergetics of hydrogenotrophic sulfate reducers is more advantageous than methanogenesis (Appendix 1) and they might outcompete or almost eliminate methanogens (Lin et al., 2006). Mars is a sulfur-rich planet (King and McLennan, 2010) and sulfates have been reported at its surface (Murchie et al., 2009; McLennan and Grotzinger, 2008; Kounaves et al., 2010). Significantly, hydrogen and sulfates can be generated concurrently through the radiolysis of water within rocks containing pyrite minerals (Lefticariu et al., 2010), turning autotrophic sulfate reduction into a plausible metabolic pathway in the Martian subsurface.

2) Do the environmental conditions of DCBF, SGL and DHAB shape the significance of subsurface methanogens and their metabolic pathways?

Neither subzero temperatures nor high salinity necessarily pose an insurmountable barrier for terrestrial methanogens (or other microbes). Methylotrophic pathways often dominate in environments characterized by high ionic strength, high sulfate concentrations, low temperatures, and neutral to low pH waters. Conversely, hydrogenotrophs prefer moderate-high temperatures, low ionic strength, low sulfates and high pH. However, that separation is not absolute. Indeed, strict hydrogentrophs have also been reported at low temperature and high salinity.

The comparison between environments indicates that DCBF sites are more challenging habitats for methanogens than DHAB sites. Two key observations support this conclusion: 1) Methanogens are ubiquitous at DHBA but not at DCBF sites; 2) the ionic strength threshold limit for methanogens is around 2 M at DCBF sites whereas at DHAB sites it reaches approximately 9 M.

3) Is the almost “single species” methanogenic ecosystem reported at Lidy Hot Springs representative for DCBF, SGL and DHAB systems?

It is well established that the terrestrial methanogens often proliferate in complex syntrophic associations with bacteria and probably with viruses (Evans et al., 2019; Wang et al., 2022). This evidence has even led to the hypothesis that eukaryotes evolved precisely from a syntrophy based on methanogens. Indeed, methanogens in subsurface enviroments belong to a taxonomically and metabolically complex community typically dominated by bacteria. Therefore, the monospecific methanogenic microbial community reported at Lidy Hot Springs does not represent a typical microbial community from subsurface terrestrial habitats. Consequently, if terrestrial-like methanogens were to exist in the Martian subsurface they would likely interact with a range of diverse organisms rather than forming single-species ecosystems.

4) How do the answers to the previous questions impact on the viability of terrestrial methanogens in Mars's subsurface?

If the optimal habitable Martian anoxic subsurface is expected to be a cold (subzero) hypersaline water body, attention should preferentially be focused on more versatile methanogens such as the Methanosarcinaceae-like and the Methanomicrobiaceae-like families rather than the Methanobacteriaceae reported in hot waters like Lidy Hot Springs. The Methanosarcinaceae-like organisms are candidates because they are able to feed on methyl-compounds, $H_2/CO_2$ and acetate (Buan et al., 2018) and tolerate high sulfate concentrations, which might represent an advantage on a sulfur-rich planet such as Mars (King and McLennan, 2010). Nonetheless, both Methanosarcinaceae and Methanomicrobiaceae are likely able to proliferate under these extreme conditions by establishing syntropic associations with fermenters or acetate oxidizers (Pan et al. 2021), and thus they should not establish "one-species ecosystems". Our literature survey revealed that the availability of $H_2$ and small organics (such as formate and acetate) does not guarantee the proliferation of methanogens at DCBF sites. This situation might be extrapolated to Mars's subsurface if sites of active production of $H_2$, formate and acetate exist. Moreover, the apparent reduction of the salinity maxima threshold reported at DCBF sites when compared to DHBA sites might have severe implications for the habitability of putative deep Martian rock-fractured fluids. At DCBF sites, other constraints besides salinity might limit methanogens, e.g. competition with other metabolisms for limiting resources (i.e. $H_2$), an excess of strong oxidants produced by water radiolysis, or physical factors such as porosity (Tanikawa et al., 2018). Importantly, the in-situ salinity tolerance threshold is based on information from NaCl- and $Ca/MgCl_2$-rich brines. More "hostile" (in a biological context) and oxidative salts, such as perchlorates, have been identified on Mars (Clark and Kounaves, 2016). On Earth, natural perchlorate brines are limited to a few hyper-arid habitats like the Atacama Desert (Ericksen, 1981), and knowledge regarding the tolerance of terrestrial biota (including methanogens) to these conditions is largely derived from experimental bioassays. These experiments indicate that it is unlikely that terrestrial methanogens proliferate in perchlorate solutions with an ionic strength greater than 0.03 M (Shcherbakova et al., 2015; Serrano et al., 2019). Therefore, the habitability of a Martian subsurface brine in fractured rocks would be drastically reduced if perchlorates accumulated in the Martian subsurface.

5) What environments in Mars's subsurface could support terrestrial-like methanogens? What methanogenic pathways might prevail?

By combining the available criteria and data on water availability and the abundance and spatial heterogeneity of HPE, we were able to identify Martian regions that might host deep water and a relatively high concentration of radiogenic elements. Our most robust target area is the southern Acidalia Planitia at mid latitude (as low as 35°N).

The target site overlaps with a large area with a high density of cones and domes (Mills et al., 2024) that have been interpreted as mud volcano-like structures (Farrand et al., 2005; Broz et al., 2023). If this interpretation is correct, it supports the presence of fluid (water-saturated mud and gases) upwelling in a not-yet identified period during the Amazonian (Broz et al., 2023) and provides clues about the Martian subsurface hydrosphere (Allen et al., 2009) (Figure 4b). Additionally, the target region almost overlaps with and area with highest density of  recurrent slope lineae (RLS) formations in the entire northern lowlands (Stillmanm et al., 2016) (Figure 4b). Although a consensus about mechanisms giving rise to RLS is lacking (McEwen et al., 2018), one of the hypotheses supports the presence of water/brine seepage (Rummel et al., 2014). Finally, relatively close to the target region is the McLaughlin Crater, which has exposed clay and carbonate deposits together with signs of groundwater activity (Michalski et al., 2013) (Figure 4b). Ultimately, combining all these observations with our analysis led to the conclusion that subsurface of the southern Acidalia Planitia could hold significant astrobiological potential, thus reinforcing Broz et al.'s (2023) recommendation that this region should be prioritized for future missions.

In the target area the subsurface temperature profiles have been estimated up to a depth of 9 km. Crucially, according to the InSight mission, salts are apparently scarce in the subsurface at low latitudes. Extrapolating this finding to the Acidalia Planitia implies that the most probable temperature at which porous ice melts would be around 0ºC. The depth at which ice melts and methanogens can proliferate depends on the depth of the ice/water datum. A hypothetical subsurface with an ice datum near the surface might be inhabitable because the zero-degree geotherm might be located at approximately 9 km depth. At this depth, the pressure-induced collapse of porosity is a significant possibility. Although this scenario needs further validation, this possibility allows speculation that the lower depth limit of a putative habitable subsurface on Mars might be determined by a physical factor rather than thermal or geochemical factors.

The depth of the zero-degree geotherm can decrease to 4.3 km in depth if the upper 2 km of insulating regolithic layers are dry. If this condition is met, then terrestrial-like bacteria-Methanosarcinaceae (and Methanomicrobioaceae) syntrophic clusters can proliferate at those depths with temperatures in the range 0–10ºC. In contrast, more thermophilic methanogenic genera, such the hydrogenotrophic Methanobactericeae, might be constrained preferentially at the highest depths provided that the putative collapse of porosity is not reached.

6) Conclusions and perspectives

Two main conclusions can be drawn from the present study:

First, methanogens are significant, but not ubiquitous, in the microbial ecosystems thriving in the subsurface ecosystems believed to be terrestrial analogs of the Martian subsurface. Thus, although methanogens feed on simple molecules and their metabolism can be described by straightforward redox equations – making them ideal models for astrobiologists, – studies of terrestrial deep temperate-hot crystalline fractures are revealing that these habitats are challenging for methanogens. Consequently, the cold Martian subsurface is likely an even harsher habitat for putative terrestrial-like methanogens.

Beyond the need for water, appropriate environmental conditions, and adequate energetic and carbon resources, the proliferation of methanogens also depends on complex ecological constraints. Indeed, methanogens establish ecological and energetic interactions with organisms with diverse metabolisms, suggesting that the conjecture that they might form mono-specific communities does not fit with what is regularly observed in terrestrial subsurface ecosystems. So, if terrestrial-like methanogens were to thrive in Mars's subsurface, they would most likely be members of a complex and diverse ecosystem.

Second conclusion: the subsurface of the southern of Acidalia Planitia is a putative target region for hosting cold-adapted Methanosarcinaceae-like and Methanomicrobiaceae-like methanogens (if they can associate with bacteria-like organisms). In this region, the radiogenic heat-producing elements are at the highest abundance and subsurface water is likely. This not only brings the 0 ºC temperature as shallow as 4.3-8.8 km below the surface but also favors water radiolysis, which potentially supplies the energetic resources required for these hypothetical methanogens.

Our knowledge of Mars's subsurface is advancing thanks to orbiters, landers and rovers, but critical gaps exist. To make substantial progress in identifying habitable niches in the subsurface of Mars, it will be essential to elucidate the availability of inorganic carbon in the subsurface, and to determine more accurately the depth at which water is located and the porosity/fracturing of the regolith, as these factors directly affect the thermal gradients and the efficiency of water radiolysis.

As a result, both our analysis (which builds upon recent advances in understanding Mars's subsurface) and previous research more focused on its surface converge in identifying the southern of Acidalia Planitia as a promising target area for future missions in the search for extant life in Mars' subsurface.

Acknowledgments

This study was funded by MCIN/AEI/10.13039/501100011033 (PID2021-123735OB-C22) and by "ERFD A way of making Europe". AB is a member of the SGR976. OF acknowledges financial support from the Departament de Recerca i Universitats of Generalitat de Catalunya through grant 2021SGR00679. OF was (partially) supported by the Spanish MICIN/AEI/10.13039/501100011033 and by "ERDF A way of making Europe" by the "European Union" through grant PID2021-125627OB-C31, and through the "Center of Excellence María de Maeztu 2020-2023" award to the ICCUB (CEX2019-000918-M). DGC was funded via the projects AGAUR 2021 SGR00410 and PID2022-139943NB-I00. The authors thank Gareth Morgan and the SWIM team for providing ancillary data for the ice contingency maps, Laura Parro for providing data on Mars modeled surface heat flows, Mackenzie Mill for providing location of pitted cones, Martin Schimmel for the enlightening conversations about the InSight mission and Alfonso Mota for comments on a previous version of this manuscript. We thank S.T.

Cady and the anonymous referee for comments and suggestions that strongly helped to improve our initial submission.

Contributions

A.B. conceived the study, collected the information from the literature and wrote the original draft followed by review and editing. O.F., R.B. and D.G.C. collected the information from the literature, wrote and supervised the manuscript, made figures and reviewed and edited the final version.

Conflict of interest.

None

# Potential habitability of present-day Mars subsurface for terrestrial-like methanogens

**Butturini A.[1], Benaiges-Fernandez R.[2], Fors O.[3], Garcia-Castellanos D.[4]**

5) Departament de Biologia Evolutiva, Ecologia y Ciències Ambientals, Universitat de Barcelona, Diagonal 643, Barcelona, Catalonia 08028, Spain
6) Departament de Genètica, Microbiologia i Estadística, Universitat de Barcelona, Diagonal 643, Barcelona, Catalonia 08028, Spain
7) Departament de Física Quàntica i Astrofísica, Institut de Ciències del Cosmos (ICCUB), Universitat de Barcelona, IEEC-UB, Martí i Franquès 1, 08028 Barcelona, Spain
8) Geosciences Barcelona (GEO3BCN), Consejo Superior de Investigaciones Científicas (CSIC), Barcelona, Spain

Corresponding author: abutturini@ub.edu

# Supplementary material

## Appendix 1

### Methanogenic pathways and resource competition with other organisms

Examples of the three Archaea-mediated methanogenic pathways with the description of the metabolic reactions and their standard free Gibbs energies ($\Delta G^o$).

d) Hydrogenotrophs, which extract energy from the $CO_2$, $HCO_3^-$ or $HCOO^-$ ($e^-$ acceptor) and $H_2$($e^-$ donor) couple following the redox:

$$+HCO_3^- + H^+ + 4H_{2(aq)} \rightarrow CH_{4(aq)} + 3H_2O \qquad \Delta G^o = -57.5\ kJ/molH_2 \qquad (1)$$

$$HCOO^-_{(aq)} + 3H_{2(aq)} \rightarrow CH_{4(aq)} + H_2O + OH^- \qquad \Delta G^o = -93\ kJ/molH_2 \qquad (2)$$

e) Acetatotrophs, which extract energy from degradation of acetate:

$$CH_3COO^-_{(aq)} + H^+ \rightarrow CO_{2\,(aq)} + CH_{4(aq)} \qquad \Delta G^o = -51.1 kJ/mol\ \ acetate \qquad (3)$$

f) Methyltrophs, which extract energy from degradation of C1 organic compounds (formate, methanol, methylamines among others, see Kurt et al., 2020 for more examples). Methyltrophs can use, or not, $H_2$ as electron donor. The following examples are with methanol:

$$4CH_3OH \rightarrow 3CH_4 + CO_2 + H_2O \qquad \Delta G^o = -65\ kJ/molCH_3OH \qquad (4)$$

$$CH_3OH + H_2 \rightarrow CH_4 + H_2O \qquad \Delta G^o = -113\ kJ/molH_2 \qquad (5)$$

$\Delta G^o$ of all reactions are estimated under standards conditions according to Amend and Shock (2001).

In addition, some methanogens can use more complex organics such as alcohols and methox groups from several aromatic compounds by-product of lignin degradation (Evans et al., 2019).

Some methanogenic groups are metabolically versatile. For instance, many hydrogenotrophs replace the $H_2$ and $CO_2$ with organics (i.e. formate, methanol) as electron donors (Long et al., 2017) and carbon source (Brazelton et al., 2017). Others grow on CO instead of $CO_2$ (Daniels et al., 1977); The genus of methanosarcina are highly metabolically diverse and can be hydrogenotrophs, acetatotrophs and metylotrophs (Maeder et al., 2006; Liu and Whitman, 2008). The most extreme example of that metabolic flexibility is provided by several methanogens strains that can oxidize methane (Zehnder and Brock, 1979; Loyd et al., 2011; Timmers et al., 2017; Bhattarai et al. 2019). This example reveals a thigh connection between methanogens and methanotrophs.

Finally, methane release as secondary metabolite has been reported in cyanobacteria, fungi, algae and plants in presence of reactive oxygen species, iron and methyl groups (Ernst et al., 2022).

Besides methanogenesis, hydrogen is the crucial electron donor for several other relevant metabolisms: oxygen, nitrate, iron, sulfate-reductions and homoacetogenesis. In consequence, organisms compete for hydrogen and metabolisms thermodynamically more advantageous should outcompete the less advantageous.

Under anoxic conditions, bioenergetics reveal that hydrogenotrophic methanogens (reaction 1) compete with bacteria sulfate reducers (reaction 6) and bacterial homoacetatogens (reaction 7).

$$4H_{2(aq)}+SO_4^{2-}{}_{(aq)} +2H^+\rightarrow 4H_2O +H_2S \qquad \Delta G^o=-76 \text{ kJ/molH}_2 \qquad (6)$$

$$4H_{2(aq)}+2HCO_3^-+H^+ \rightarrow CH_3COO^-{}_{(aq)} +4H_2O \qquad \Delta G^o=-54 \text{ kJ/molH}_2 \qquad (7)$$

According to the $\Delta G^o$ values of reactions 1, 6 and 7, the methanogens hydrogenotrophs should be outcompeted by sulfate reducers but, in their turn, they should exclude the homoacetatogens.

However, in some situations the $\Delta G^o$ value can be misleading and, if information of chemical composition of the media is available, the standard $\Delta G^o$ value must be replaced by the actual Gibbs energy of the reaction, $\Delta G_r$ (Amend and de LaRowe, 2019). Under specific environmental conditions, differences between $\Delta G_r$ of reactions 1 and 5 can change substantially, or even can be inverted (Trutschel et al., 2022).

Thermodynamic estimations have important limitations: reactions are assumed to occur under equilibrium conditions; do not consider any kind of limitation in organisms' activities/grow (for instance the availability of trace nutrients); do not consider the energy required to transport the (charged)resources/residues/protons across the cell membranes; do not consider the energetic cost to counteract the osmotic, chaotropic stresses; do not consider the flexibility of organisms to feed on other resources. Indeed, regardless to thermodynamic, methanogens can proliferate together to sulfate reducers (Ozuolmez et al., 2015; Ma et al., 2017) or be totally inactive albeit sufficient energy is available (Telling et al., 2018).

The same occurs with homoacetatogens. The thermodynamic difference between homoacetatogenesis and methanogens is of a few kJ. Indeed, coexistence among the two metabolisms has been reported especially under low temperatures (<15ºC) and unlimited $H_2$ availability (Tsapekos et al., 2022; Kotsyurbenko et al., 2001).

In summary, bioenergetic calculations are a critical initial step to speculate about which metabolisms have the potential to occur under specific environmental conditions. However, these calculations need to be completed with direct observations and molecular or genetic approaches.

## Appendix 2

**Survival of methanogens under putative Martian environmental conditions. Experimental tests**

The following list reports experimental studies that tested the survival of methanogen strains (i.e. monoculture) under Mars environmental conditions. Studies are sorted according to the tested main variable.

*Temperature:*

*Pressure:*

Minerals and carbon sources & Martian-like regolith:

**Appendix 3**

**DCBF, SGL and DHAB habitats in the context of Mars' subsurface habitability.**

The following table ST3 identifies the main Pro and Cons of DCBF, SGL and DHAB habitats in a context of Mars' subsurface habitability.

| Habitat | Pros | Cons |
|---|---|---|
| *Crystalline-bedrock fracture habitats (DCBF)* | Putative analog of Martian subsurface crust. Host ecosystems that potentially have been isolated from processes occurring on Earth' surface during long period. Appropriate habitats to provide abiotic reducing power ($H_2$ and other small size organic molecules) for methanogens. | Ecosystems that proliferate at temperatures that are probably excessively high for those expected in a typical Martian subsurface lithosphere. Salinity is typically low in most cases. Yet, hypersaline brines are reported in ancient aquifers (Telling et al., 2015). |
| *Sub-endoglacier lentic waters and brines (SGL)* | Terrestrial analogs of subglacial water bodies detected below Martian southern ice cap. Ecosystems that proliferate at temperatures more similar to those expected at Mars subsurface. Some SGL fluids accumulate high salinities indispensable to have liquid water at below-zero temperatures. | Some of SGL studied to date are tightly connected to Earth' surface processes: host photosynthetic-oxygenic and aerobic organisms (Karr et al., 2003); are in contact with sediments rich in organic matter accumulated during interglacial periods (Wadham et al., 2008; Tung et al. 2005). Abiotic production of $H_2$ has been reported from ex-situ experiments (Telling et al., 2015) but in-situ production $H_2$ has not been studied in detail. |
| *Anoxic sub-sea hypersaline waters (DHAB)* | Putative analogs of expected Martian deep hypersaline water bodies. Biota that thrives under salinity/ionic strength conditions extremely high. DHAB originates from dissolution of evaporitic deposit as consequence of infiltration of seawater. Salts are indispensable to have liquid water at below-zero temperatures and are expected to be relevant on Mars subsurface. In situ abiotic leaking of $H_2$ and small organics has been reported at DHAB in contact with hot springs. | Ecosystems connected to Earth' biotic processes occurring at surface.Metabolisms at DHBA are partially fueled by organic matter that settled from overlying oxygenated water column (Yakimov et al., 2013). For most sites, in situ abiotic production of $H_2$ and small organics has not been studied. |

ST3

**Appendix 4**

**Material and Methods**

Microbial database:

We collected 94 papers that summarized the information from 79 sites. Methanogens are described at level of “family”.

When, at a study site, methanogens are not described or not detected, in our database we indicate “not reported”. Genomic techniques have limitations, especially for detecting strains at low relative abundances. Therefore, no detection does not necessarily mean that methanogens are missing (Ward et al., 2004).

In our review the Archaea candidate division MSBL-1(Mediterranean Sea Brine Lakes-1, member of Thermoplamata) is not considered a methanogen because its metabolism is under scrutiny and some authors proposed that MSBL-1 is mixotroph (i.e. fermenter and autotrophic; Mwirichia et al., 2016).

Besides the DCBF, DHAB and SGL sites, additional terrestrial subsurface habitats that host complex subsurface microbial communities are the deep continental sedimentary setting, deep sub-sea sediments, deep sea hydrothermal systems and, springs from continental groundwaters. Nevertheless, information from these systems has been excluded from the data analysis. Below arguments that motivate our decision are provided:

I. Continental sedimentary deep settings and sub-sea sediments because can accumulate organic matter originated (in the past) by the photosynthetic marine and terrestrial ecosystems (Pinero et al., 2013; Arndt et al., 2013) and because oxygen can penetrate the sediments below the sea floor (Røy et al., 2012). Therefore, these systems are probably excessively connected to the oxidant biosphere.

II. Deep sea hydrothermal systems, because these systems exchange water with overlying oxygenated water column (Stewart et al., 2019). Evidence of hot waters at present Mars subsurface is missing. Last but not least important, on present-day Mars the sea and associated deep-sea habitats are absent.

III. Surficial springs from continental groundwaters, because are in contact with atmospheric oxygen that is prejudicial for methanogens (for instance, the alkaline springs at Voltri massif, Tablelands, the Blood falls in Antarctic, hot springs at Bad Gastein, Austria, hypersaline springs at Canadian high Arctic or cold sulfidic springs from Regesburg, Germany).

Salinity proxy:

Ionic strength (I) is a measure of the concentration of electric charged solutes of a solution, and it considers both the molar concentration (*c*) and charge (*z*) of ions ($I = \frac{\sum_{i=1}^{n} c\, z^2}{2}$). I values are provided in several DHAB sites. But for most DCBF and SGLs sites I is not provided and has been estimated according to the concentrations of $Cl^-$ and $SO_4^{2-}$ (when available) and if these ions come from dissolution of NaCl and $CaSO_4$ salts respectively.

At the subglacial lake Vanda, (Saxton et al. 2021), brines are from dissolution of $CaCl_2$. In this specific case, I value is estimated with concentrations of $Ca^{2+}$ and $Cl^-$.

When only concentration of $Cl^-$ is available, we assumed that all I originates by dissolution of NaCl and therefore I coincides with the concentration of $Cl^-$.

Mars subsurface:

*Water/ice:*

Information about water/ice presence at Mars subsurface's originates from observations performed locally with in-situ instrumentation and globally and with remote sensing. Sources of local scale observations come from in-situ radar profiles (0-140 m depth) performed by the Zhurong rover at the southern marginal area of Utopia basin (25.1N, 109.9E) (Li et al, 2022); seismic data of the upper crust (0-10 km) from InSight at Elysum Planitia (4.5N, 135.6E) (Manga and Wright, 2021); direct imaging of surface at Phoenix (68.2 N, 234.3 E) and Viking 2 (47.6N; 135E) landing sites.

On a more global scale, we feed on the synthesis provided by Morgan et al., (2021). These authors synthetize and visualize the information obtained with several remote-sensing techniques through global maps showing the likelihood of ice at surface (<1 m depth), very shallow subsurface (1-5 m depth) and more buried cryosphere (>5 m depth). The probability of ice is visualized in terms of contingency maps. We adopted the information provided by the contingency map for depth greater than 5 m which integrated the information provided by SHARAD radar and geomorphological evidences (Putzig et al., 2023). This information is available at https://swim.psi.edu/.

*Energy sources for methanogens:*

The four mechanisms that might support $H_2$ accumulation at the Martian subsurface/regolith are:

i) Photo production of $H_2$ and its diffusion from atmosphere to subsurface porous and dry regolith is modeled by Sholes et al. (2019).

ii) Rock friction and crushing of rocks and minerals (hydrsted and/or in contact with water) after seismic movements (McMahon et al. 2016).

iii) Water-rock interactions as serpentinization within crustal fractures explored by Amador et al. (2018) implementing the information provided by the Compact Reconnaissance Imaging Spectrometer (CRISM).

iv) Water radiolysis is fueled by crustal radiogenic elements.

Putative significance of water radiolysis is explored analyzing the abundance of crustal radiogenic elements. The Mars Odyssey Mission Gamma Ray Spectrometer (GRS) provides data of availability of K and Th on Mars surface across all longitudes and from 60ºS to 60ºN (Boynton et al., 2007). Data set is available at https://ode.rsl.wustl.edu/mars/pagehelp/Content/Missions_Instruments/Mars_Odyssey/

GRS/IDR/ELEMTS.htm. This information allows us to estimate the abundance and spatial distribution of radiogenic elements ($^{232}$Th, $^{40}$K, U which is inferred adopting a ratio Th/U=3.8, Hahn et al., 2011).

*Geothermal gradients:*

Temperatures at Mars surface are modeled with the Mars climate database (Millour et al., 2018). A global map of modeled present-day surface heat flow is provided by Parro et al. (2017).

Regarding the subsurface geology of Martian upper crust, the information about number of layers and their thickness, densities and thermal conductivities is collected from several reports summarized at Table ST4. Of special interests are the subsurface geological structure derived from the interpretation of seismic waves recorded by the Insight from marsquakes and impact events with epicenters at the northern lowlands (Knapmeyer-Endrun et al., 2021; Xu et al., 2023; Shi et al., 2023).

Table ST4 puts emphasis on the ice/water content. Pure ice has a thermal conductivity of around 2.2 W/m/K and thermal conductivity of porous ice-saturated material is typically higher than its dry contour part. For instance: thermal conductivity of saturated icy Mars upper regolith analogs with 40% of porosity is estimated around 0.9-1.2 W/m/k (Siegler et al., 2012) in contrast the thermal conductivity of its dry counterpart is likely lower than 0.08 W/m/k (Grott et al., 2021).

| **Geologic structure** | **Depth range (km)** | **Density [kg m$^{-3}$]** | **Thermal conductivity [W m$^{-1}$ K$^{-1}$]** | | **References** |
|---|---|---|---|---|---|
| Upper regolith | Unconstrained (few cm as much) | 1210 | 0.039-0.08 (dry) | 0.9-1.2 (wet) | Zent et al., 2010; Grott et al., 2021; Siegler et al., 2012. |
| Highly fractured megaregolith | 0-2 | 1700-2100 | 0.8 (dry) | 2.4 (wet) | Egea-González et al., 2021; Shi et al., 2023 |
| Fractured basalt | 2-9 | 2750 | 2.5<br>488.2/T+0.47 | | Halbert & Parnell, 2022; Egea-González et al., 2021 |

| ice/water saturated | | | | Kilburn et al.,2022; Xu et al., 2023; Clifford et al., 2010 |
|---|---|---|---|---|
| Unfractured basaltic crust | 9-20 | 2900 | 3.6-4.9 $10^{-3}$ T +0.61 $10^{-5}$ $T^2$- 2.58 $10^{-9}$ $T^3$ ($T>150$ K) | Egea-González et al., 2021; Schimmel et al., 2021; Halbert & Parnell, 2022; Zoth and Haenel (1988) |

ST4. Estimates of thickness, densities and thermal conductivities of four geological structures expected for Mars upper crust according to the literature.

**Appendix 5**

**The biological data set**

The following appendix includes the files “AlldataSET.xlsx” and “Bibliographic_sources.xlsx”.

File “AlldataSET.xlsx” reports all data from DCBF, DHBA and SGL sites compiled in this study. This file comprises the environmental and microbial information available at each location. Environmental information includes the geological setting, age of aquifer, sample substrate (i.e. water or rock), depth, temperature and a set of biogeochemical descriptors (pH, ionic strength, $Cl^-$, $SO_4^{2-}$, dissolved organic carbon –DOC-, dissolved inorganic carbon-DIC, $H_2$, and $CH_4$). The microbial information includes the name of reported methanogen family, and its expected methanogenic pathway according to the information provided by authors and that available at the PhyMet2 database (Michał et al. 2018). A full description of bacterial microbiome at each site is beyond the objective of this study. However, the most relevant bacteria/archaea genera detected/reported at each site are mentioned at the last column of that table.

$H_2$ and $CH_4$ values reported in the file are, in most of cases, the concentration of dissolved gas concentration (expressed in $\mu$M). However, in some cases the values describe the volume in the gas phase (% vol).

File “Bibliographic_sources.xlsx” reports the references used to compile the file “AlldataSET.xlsx”.

**Appendix 6**

**Availability of acetate and formate and Methanogenesis rates reported at SGL, DHBA and DCBF sites.**

The following table ST6 reports the concentration of acetate and formate values at several subsurface terrestrial habitats together to estimate methanogenesis rates when available.

| | | | | Methanogenesis Rates (μM/d) | | | | |
|---|---|---|---|---|---|---|---|---|
| **HABITAT** | **SITE** | **Acetate (μM)** | **Formate (μM)** | **Hyrogen otroph** | **Acetatrop h** | **Mthylotro ph** | **TOTAL** | **SOURCE** |
| SGL | Lake Untersee (water column) | 5.5-11 | NA | 0.02-0.44 | $<0.08\ 10^{-3}$ | NA | <0. 45 | Wand et al. (2006) |
| SGL | Lake Untersee (sediments) | 180-310 | NA | 0.26-0.87* | 0.005-0.02* | NA | <0.9* | |
| SGL | Lake Vanda | NA | NA | $0\text{-}10^{-6}$ | $<0.1\ 10^{-6}$ | $0\text{-}40\ 10^{-6}$ | $<41\ 10^{-6}$ | Saxton et al. (2021) |
| SGL | Lake Whillans | 1.3 | 1.2 | NA | NA | NA | NA | Christner et al., 2014 |
| DHAB | Atlante Brine (Med) | NA | NA | NA | NA | NA | 16.5 | van der Wielen et al. (2005) |
| DHAB | Bannock Brine (Med) | NA | NA | NA | NA | NA | 4.2 | van der Wielen et al. (2005) |
| DHAB | Urania Brine (Med) | NA | NA | NA | NA | NA | 85.8 | van der Wielen et al. (2005) |
| DHAB | Discovery Brine (MEd) | NA | NA | NA | NA | NA | 2.7 | van der Wielen et al. (2005) |
| DHAB | Lake Medee | 132-539 | NA | NA | NA | NA | 0.5-2 | Yakimov et al (2013) |
| DHAB | Gulf of Mexico Brine Pool | 2.6-68 | 1.7-5.7 | $<10^{-6}$ | $<0.27\ 10^{-3}$ | NA | $<0.27\ 10^{-3}$ | Joye et al. (2009) |
| TaDHAB | Gulf of Mexico Volcano Mud | 1.7-57 | 8.0-34 | <0.02 | <0.12 | NA | <0.12 | |
| DCBF | Driefontein | 35 | 3.5 | NA | NA | NA | NA | Moser et al., 2005 |
| DCBF | Driefontein (DR5IPC) | NA | NA | $4.1\ 10^{-6}$ | NA | NA | NA | |
| DCBF | Beatrix (BE326Bh2) | <1.6 | <8.4 | NA | NA | NA | NA | Simkus et al., (2016). |
| DCBF | TauTona (TT109Bh2) | 0.16 | 1 | $1.8\ 10^{-05}$ | NA | NA | NA | |
| DCBF | Kloof (KL445) | 28 | 0.9 | NA | NA | NA | NA | |
| DCBF | Kidd Kreek | 1400-1900 | 500-1000 | NA | NA | NA | NA | Lollar et al., 2021 |
| DCBF | Thompson Mine | 53-496 | BDL | 0.1** | NA | NA | NA | Telling et al. 2018 |
| DCBF | Samail Ophiolite (NSHQ14) | 1.2 | 1.7 | 0.02 | NA | NA | NA | Rempfert et al. (2017). |
| DCBF | Samail Ophiolite (WAB188) | 3.8 | 1 | 0.03-2.2 | NA | NA | NA | Fones et al. (2019) |
| DCBF | Mizunami Underground | <3.3 | <1.8 | NA | NA | NA | NA | Fukuda et al. (2010) |
| DCBF | Mponeng | <36 | <8.9 | NA | NA | NA | NA | Lin et al., (2006) |
| DCBF | Lidy Hot | <1 | <1 | NA | NA | NA | NA | Chapelle et al., (2002) |
| DCBF | Coast Range (QV1.1) | 10.2 | <1.4 | NA | NA | NA | NA | Twing, et al., (2017) |
| DCBF | Coast Range (CSW1.1) | 71 | 15.8 | NA | NA | NA | NA | |
| DCBF | Lupin Au Mine (890-188) | 0.02 | 0.27 | NA | NA | NA | NA | Onstott, (2009) |
| DCBF | Lupin Au Mine (1130-192) | 0.03 | 8.8 | NA | NA | NA | NA | |
| DCBF | Iberian Pyrite Belt | 800 | 290 | NA | NA | NA | NA | Puente-Sánchez et al., 2014 |

Table ST6

**Appendix 7**

**Metabolisms heterogeneity at terrestrial subsurface ecosystems**

The following Figure SF7 syntheses the main metabolic pathways reported in terrestrial depth biosphere and how they interact with each other (yellow arrows). In blue are the geological processes.  All organisms produce biomass. The fraction of biomass that is not respired into $CO_2$ converges into the "Organic matter" pool.

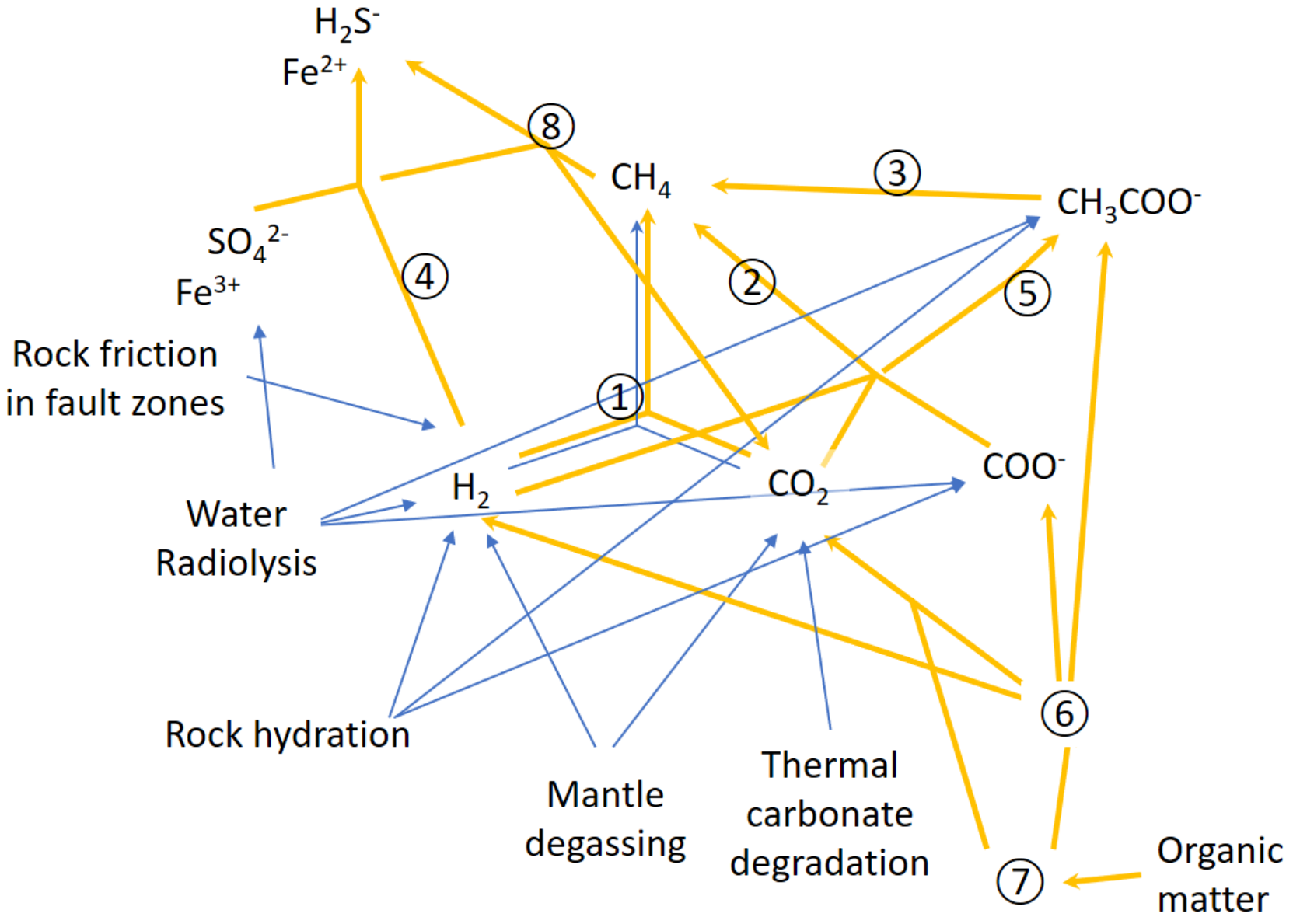

① Hydrogenotrophic methanogenesis
② Methylotrophic methanogenesis
③ Acetatotrophic methanogenesis
④ Autotrophic quimiolito reducers
⑤ Autotrophic homoaceteatogenesis
⑥ Fermenters
⑦ Anaerobic respirations
⑧ Methanotrophy

Figure SF7.